\documentclass[12pt]{article}
\usepackage{epsfig}
\usepackage{amsfonts}
\begin{document}
\begin{titlepage}
\begin{center}

{\Large Non-additivity of Tsallis entropies and fluctuations of temperature}

\vspace{2.cm} {\bf Christian Beck}\footnote{
permanent address: School of Mathematical Sciences, Queen Mary,
University of London, Mile End Road, London E1 4NS.}

\vspace{0.5cm}

Isaac Newton Institute for Mathematical Sciences,
University of Cambridge, 20 Clarkson Road, Cambridge CB3 0EH, UK

\vspace{5cm}

\end{center}

\abstract{We show that the
non-additivity relation of the Tsallis entropies 
in nonextensive statistical mechanics has 
a simple physical interpretation for systems 
with
fluctuating temperature or fluctuating energy dissipation rate.
We also show that there is a distinguished
dependence of the entropic index $q$ on the spatial scale
that makes the Tsallis entropies quasi-additive. Quasi-additivity
implies that
$q$ is a strictly monotonously decreasing function of the
spatial scale $r$, as indeed observed in various
experiments.}

\vspace{1.3cm}

\end{titlepage}

The formalism of nonextensive statistical mechanics 
has been developed over the past 13 years
as a powerful and beautiful generalization
of ordinary statistical mechanics \cite{tsa1}--\cite{web}. 
It is based on the extremization of the
Tsallis entropies
\begin{equation}
S_q= \frac{1}{q-1} \left( 1- \sum_i p_i^q \right)
\end{equation}
subject to suitable constraints. Here the $p_i$ are
the probabilities of the physical microstates, and
$q$ is the entropic index.
The Tsallis entropies
reduce to the Boltzmann Gibbs (or Shannon) entropy
$S_1=-\sum_i p_i\log p_i$ for $q\to 1$, hence ordinary
statistical mechanics is contained as a special case
in the generalized formalism.
There is
growing experimental
evidence that $q\not= 1$ yields a correct decription of
many complex physical phenomena, including hydrodynamic turbulence 
\cite{hydro}--\cite{ari},
scattering processes in particle physics \cite{e+-1,e+-2}, and
self-gravitating systems in astrophysics \cite{plastino,astro2},
to name just a few.
A complete list of references 
can be found in \cite{web}.

The Tsallis entropies are non-extensive. Given
two independent subsystems I and II with probabilities
$p_i^I$ and $p_i^{II}$, respectively, the entropy of the composed system
I+II (with probabilities $p_{ij}^{I+II}=p_i^Ip_j^{II}$)
satisfies
\begin{equation}
S_q^{I+ II}=S_q^{I}+S_q^{II}-(q-1)S_q^{I}S_q^{II} \label{nonad}.
\end{equation}
Hence there is additivity 
for $q=1$ only. This property has
occasionally lead to unjustified prejudices, using 
arguments of the type `entropy must be extensive'.
In the following, however, we will see
that the non-additivity property is not at all a `bad'
property,
but rather a natural and physically consistent property
for those types of systems where nonextensive statistical mechanics
is expected to work.

Broadly speaking,
the formalism with $q\not=1$ has so far been observed to
be relevant for two different classes of systems.
One class is systems with long-range interactions 
(e.g.~\cite{plastino,astro2,ruffo}), the other one is
systems with fluctuations of temperature or of energy dissipation 
(e.g.~\cite{BLS,e+-1,e+-2,wilk}). 
For the first class of systems,
the concept of {\em independent} subsystems does not make any sense,
since all subsystems are interacting. Hence there is no
contradiction with the non-additivity of entropy for
independent subsystems since independent subsystems do no exist
for these systems
(by definition of the long-range interaction). 

Let us
thus concentrate on the other class, systems with fluctuations.
Consider a system of ordinary statistical mechanics with Hamiltonian $H$.
Tsallis statistics with $q>1$ can arise from this ordinary Hamiltonian
if one
assumes that the temperature $\beta^{-1}$ is locally
fluctuating. 
From the integral representation of the gamma function one can easily
derive the formula \cite{beck01,wilk} 
\begin{equation}
(1+(q-1)\beta_0 H)^{-\frac{1}{q-1}}= \int_0^\infty e^{-\beta
H} f(\beta ) d\beta \label{marl},
\end{equation}
where
\begin{equation}
f (\beta) = \frac{1}{\Gamma \left( \frac{1}{q-1} \right)} \left\{
\frac{1}{(q-1)\beta_0}\right\}^{\frac{1}{q-1}} \beta^{\frac{1}{q-1}-1}
\exp\left\{-\frac{\beta}{(q-1)\beta_0} \right\} \label{fluc}
\end{equation}
is the probability density of the $\chi^2$ distribution.
The above formula is valid for arbitrary Hamiltonians $H$ und thus
of great significance.
The left-hand side of eq.~(\ref{marl})
is just the generalized Boltzmann factor emerging out of
nonextensive statistical mechanics. It can directly be
obtained by extremizing $S_q$. The right-hand side is a weighted
average over 
Boltzmann factors of ordinary statistical mechanics.
In other words, if we consider a nonequilibrim system (formally
described by a fluctuating $\beta$), then the generalized
distribution functions of nonextensive
statistical mechanics are a consequence of integrating over all
possible fluctuating inverse temperatures
$\beta$, provided $\beta$ is $\chi^2$ distributed. 

The $\chi^2$ distribution is a
universal distribution that occurs in many very common
circumstances (see any statistics handbook on this,
e.g.\ \cite{hast}). For example, it arises
if $\beta$ is the sum of squares of $n$ Gaussian
random variables, with $n=2/(q-1)$. Hence one expects Tsallis statistics
to be relevant in many applications. 
For fully developed turbulent hydrodynamic flows, where Tsallis statistics
has been observed to work very well \cite{BLS},
$\beta^{-1}$ is not the physical temperature of the flow but
a formal temperature
defined by the fluctuating energy dissipation rate times a 
time scale
of the order of the Kolmogorov time \cite{beck01}. In the application
to scattering processes in
collider experiments \cite{e+-1,e+-2}, $\beta^{-1}$ 
is a fluctuating inverse temperature
near the Hagedorn phase transition.

The constant $\beta_0$ in eq.~(\ref{fluc}) is the average of the
fluctuating $\beta$,
\begin{equation}
E(\beta ):=\int_0^\infty \beta f(\beta) d\beta =\beta_0
\end{equation}
($E$ denotes the expectation with respect to $f(\beta)$). The
deviation of $q$ from 1 can be related to the relative variance
of $\beta$. One has
\begin{equation}
q-1= \frac{E(\beta^2)-E(\beta)^2}{E(\beta)^2}.
\end{equation}

We
can now give physical meaning to eq.~(\ref{nonad}) for $q>1$.
Consider two independent subsystems I and II that
are composed into one system I+II. 
In system I (as well as II) the fluctuations of temperature $T$ are
expected to be larger than in I+II, due to the smaller size of I (or II)
as compared to I+II. Remember that generally entropy
is a measure of missing information
on the system \cite{BS}. The entropy 
$S_q^I+S_q^{II}$ is 
larger than $S_q^{I+II}$, since in the single
systems the probability distribultion of $T$ is broader, thus our
missing information on these systems is larger.
Hence there must be a negative correction term to $S_q^I+S_q^{II}$
in eq.~(\ref{nonad}).
It is physically plausible that this correction term is
proportional a) to the relative strength of temperature fluctuations
as given by $q-1$ and b) to the entropies $S_q^I$ and $S_q^{II}$
in the single systems. Hence we end up with eq.~(\ref{nonad}).

For most physical applications
the fluctuations
of $\beta$ (or the relevant value of $q$)
are observed to be dependent on the spatial scale $r$.
For example, in the turbulence application detailed measurements
of $q(r)$ have been presented in \cite{BLS}. $q$ turns out
to be a strictly monotonously decreasing function of the
distance $r$ on which the velocity differences are measured.
Similarly, for the application to $e^+e^-$ annihilation \cite{e+-2},
$q(r)$ turns out to be again a strictly monotonously descreasing
function of the scale $r$, which
in this case is given by $r\sim \hbar/E_{cm}$,
where $E_{cm}$ is the center of mass energy of the beam. Let
us now present a theoretical argument why physical systems
may like to choose a scale-dependent monotonously decreasing $q$. 

The observation is
that it is possible to make the Tsallis entropies quasi-additive
by choosing different entropic indices at different scales.
I.e., given a certain $q$ for two small independent subsystems I and II
we may choose another $q'$ for the larger, composed
system I+II such that
\begin{equation}
S_q^I +S_q^{II} =S_{q'}^{I+II}. \label{bf}
\end{equation}
We may call this property {\em quasi-additivity}. For practical
applications, $q$ is
often close to 1,
so that a perturbative expansion in $q-1$ makes sense.
One obtains
\begin{eqnarray}
\sum_i p_i^q&=& \sum_i p_i e^{(q-1)\log p_i} \nonumber \\
&=& 1+(q-1)\sum p_i \log p_i + \frac{1}{2} (q-1)^2 
\sum p_i (\log p_i)^2+ \ldots, \label{piq}
\end{eqnarray}
where the dots denote higher-order terms in $q-1$. This
yields for the sum of entropies of the two identical subsystems
\begin{eqnarray}
S_q^I+S_q^{II}=2S_q^I&=& \frac{2}{q-1} (1-\sum_ip_i^q) \nonumber \\
&=& -2\sum p_i \log p_i -(q-1) \sum p_i (\log p_i)^2 -\ldots
\end{eqnarray}
On the other hand, by squaring eq.~(\ref{piq}) one obtains
\begin{eqnarray}
\left(\sum p_i^q\right)^2&=&1+2(q-1)\sum p_i \log p_i \nonumber \\
              &+& (q-1)^2 (\sum p_i \log p_i)^2 +(q-1)^2 \sum p_i (\log p_i)^2
+ \ldots
\end{eqnarray}
and hence
the entropy of the composed system satisfies
\begin{eqnarray}
S_{q'}^{I+II}&=&\frac{1}{q'-1} \left(1-\sum_{ij}p_{ij}^{q'}\right) \nonumber \\
&=& \frac{1}{q'-1}\left( 1-(\sum_i p_i^{q'} )^2\right) \nonumber \\
&=& -2 \sum p_i \log p_i -(q'-1) \left( (\sum p_i \log p_i)^2+
\sum p_i (\log p_i)^2\right) \nonumber \\
&\,& +\ldots 
\end{eqnarray}
Quasi-additivity thus implies a relation between $q'$ and $q$,
namely
\begin{equation}
\frac{q'-1}{q-1}=\frac{\sum p_i (\log p_i)^2}{\sum p_i (\log p_i)^2+(\sum
p_i \log p_i )^2},
\end{equation}
which can be written in the simple form
\begin{equation}
\frac{q'-1}{q-1}=\left( 1+\frac{\langle B_i \rangle^2}{\langle B_i^2 
\rangle}\right)^{-1} \label{bit}.
\end{equation}
Here $B_i:=\log p_i$ is the so-called `bit number' \cite{BS}.
The negative expectation $-\langle B_i \rangle$ of the bit number
is just the ordinary Boltzmann-Gibbs (or Shannon) entropy.
We thus see that 
quasi-additive behaviour necessarily implies a change
of $q$ with system size. For $q$ close to 1, this change
is determined by eq.~(\ref{bit}),
which involves the average and variance
of the bit number,
i.e. quantities related to fluctuations of the
Shannon entropy. 
It is now clear from eq.~(\ref{bit}) that $q'-1$ (corresponding
to the larger system) is smaller than $q-1$ (corresponding to the
smaller system) for arbitrary probabilities $p_i$. In other words, $q(r)$ is a strictly
monotonously decreasing
function of the scale $r$, just as observed in the experiments.

\end{document}